\documentstyle[12pt]{article}
\begin{document}
\begin{center}
\Large{\bf Vacuum Energy and the Cosmological Constant}\\
\vspace {.1in}
\normalsize A. C. Melissinos \\
{\it Department of Physics and Astronomy, University of Rochester, Rochester, NY 
14627} 
\vspace {.1in}

11/30/01
\end{center}
\bigskip
We discuss a numerical relation between the cosmological constant
$\Lambda$ [1] and the vacuum energy arising from the Casimir effect
[2].  Many recent theories of the elementary particles invoke the
existence of extra dimensions, beyond the familiar 4-dimensional
space-time manifold.  Often these extra dimensions are compact and
have a finite range.  In a particular model [3] the range $R$ is of the 
order of
$100\;\mu$m - 1 mm.

It is interesting to consider the vacuum energy, arising from the
Casimir effect due to the bounding of the 4-d space by the extra compact
dimension(s).  In its simplest form, that of parallel plates,
the Casimir stress is attractive.  However Boyer [4] has shown that a
conducting spherical shell of radius $a$ is subject to a \underline
{repulsive} stress due to the electromagnetic zero point energy.  This
is the desired configuration if the vacuum energy is to be identified
with the cosmological constant.

Boyer's calculation leads to an energy density
\begin{equation}
    u = {0.092\over 2a} {\hbar c\over (4\pi/3)a^3} \simeq 0.01 {\hbar
c\over a^4}
\end{equation}
where $a$ is the radius of the shell.  The energy density associated
with the cosmological constant is
\begin{equation}
    \Lambda = \Omega_\Lambda\rho_c \simeq
    3 \times 10^3 \;\;\rm eV/{\rm cm}^3
    \end{equation}
where we used $\Omega_\Lambda = 0.65$ and for the critical density
$\rho_c = (3/8 \pi)(H^2_0/G_N) = 4.5 \times 10^3 \rm eV/{\rm cm}^3$ (with
$h_0 = 2/3$) [1]. Equating (1) and (2) we find
\begin{equation}
a = 21  \;\mu{\rm m}
\end{equation}

This numerical result is of the order of the range of the extra
dimensions* mentioned previously [3]. It is important to note that the
gravitational interaction has not been tested at such short
distances.  This is relavant because the ``extra dimensions'' are
invoked in order to accommodate a gauge theory of the gravitational
interaction. 

Clearly we have glossed over the issue of the
particular field that gives rise to the vacuum energy as well as on
treating more than one extra dimension.  In general such
considerations modify the result of Eq. (3) by no more than a decade. 
A proper application of the Casimir effect to cosmology can be found in
a recent paper by Brevik {\em et al} [5].

This note originated in a discussion with Dr.P.K.Williams who had also
considered such a relationship. 
\medskip

* However, the ``thickness'' of the 4-d world in that model is only $\sim 
1/M_W$ where $M_W$ is the electroweak energy scale.

\end{document}